\documentclass[12pt]{article}

\usepackage{graphicx}

\begin{document}

\title{Interbreeding conditions for explaining Neandertal DNA in living humans: the nonneutral case\footnote{\uppercase{T}his work is supported by \uppercase{FAPEMIG}.}}

\author
{Armando G. M. Neves\\
\\
\normalsize{Departamento de Matem\'atica, Universidade Federal de Minas Gerais,}\\
\normalsize{aneves@mat.ufmg.br}
}
\maketitle

\abstract{
We consider here an extension of a previous work by Neves and Serva, still unpublished, which estimates the amount of interbreeding between anatomically modern Africans and Neandertals necessary for explaining the experimental fact that 1 to 4\% of the DNA in non-African living humans is  of Neandertal origin. In that work we considered that Africans and Neandertals had the same fitness (neutral hypothesis) and Neandertal extinction was thus an event of fortune. In this work we consider that Africans had larger fitnesses. We show results for four values for the fitness difference: 1\%, 5\%, 10\% and 20\% and compare them with the corresponding neutral results. Some technical differences with respect to the neutral case appear. We conclude that even with 1\% fitness difference Neandertals extinction comes up in too small a time, so the neutral model looks more suitable for explaining the known data on occupation of some caves in Israel for a very long time, alternately by Africans and Neandertals.}

\section{Introduction}
For many years the question of the origins of the \textit{Homo sapiens} seemed to have been decided in favor of the Out of Africa model. According to this model our species descends from a small group of Africans which appeared circa 100 to 200 thousand years ago and eventually spread all over the world. In its expansion this group would have met other human groups which already lived at other parts of the world. The most well-known of these groups is Neandertals, which inhabited Europe and West Asia until circa 30 thousand years ago and then disappeared. According to the Out of Africa model, other human groups, Neandertals in particular, belonged to different species and did not contribute to the \textit{H. sapiens} gene pool. A competing theory is the Multiregional Hypothesis\cite{wolpoff}, according to which humans are the result of parallel evolution of some more or less separate regional groups, with occasional interbreeding among them making sure a unique species resulted.

Support for the Out of Africa model came from paleonthological and archaeological arguments. But the support given by genetics was decisive. In the 1980's \cite{cann,wilson} it was shown that all living humans are very similar to each other from the point of view of mitochondrial DNA (mtDNA). So similar indeed that mtDNA data are compatible with all of us being descendants of a single woman, which was then called the \textit{mitochondrial Eve}. The time and place in which the mitochondrial Eve lived can be inferred from the distribution of differences among living humans in mtDNA and from geographical correlations. According to the supporters of the Out of Africa model, if interbreeeding with other human groups, e.g. Neandertals, had occurred, then we would have to see some rather different mtDNA types among living humans. Those different types have never been experimentally found.

More recent genetic data of another kind seemed to confirm this view. By 1997 sequencing technology had advanced to the point of being able to study mtDNA extracted from a few Neandertal fossils\cite{kringsetalcell,kringsetal} and it resulted that indeed Neandertal mtDNA is rather different of the types existing among living humans. As a result of such seemingly decisive arguments, the Multiregional Hypothesis was almost completely discredited, although some statistical results favoring it could still be found\cite{templeton2005}.

Maurizio Serva\cite{serva1,serva2,serva3} and myself in collaboration with Carlos H. C. Moreira\cite{nm1,nm2,nm3} observed in independent works that the genetic arguments based on mtDNA utilized to support the Out of Africa model were not sufficient to conclude that the Multiregional Hypothesis was wrong. In order to have conclusive data we would have to resort to nuclear DNA. Very recent experimental data showed that we were right. Green et al\cite{greenetal} managed in 2010 to produce a first draft of the Neandertal nuclear genome. By comparing this genome with that of living humans from different geographic origins, they found out that Neandertals are genetically closer to present day non-Africans of either origin than to Africans. Not only this discovery provided a direct proof of interbreeding among Neandertals and our African ancestors, but it also allowed to estimate that living non-Africans carry in their cell nuclei 1 to 4 \% DNA of Neandertal origin. Moreover, the fact that all non-Africans seem to be equally closer to Neandertals suggests that the interbreeding might have occurred in the Middle East between 45 to 80 thousand years ago, when both groups occupied that region and before Africans had spread to the rest of the world. The strict Out of Africa model was definitely disproved. A second blow came still in 2010, when Reich et al\cite{reichetal} discovered that another human group previously known only through its distinctive mtDNA did also contribute to the nuclear genetic pool of living Melanesians.

The natural question posed by these discoveries is how much frequent the interbreeding should have been in order to explain the 1 to 4 \% proportion of Neandertal DNA in living non-Africans. Myself and Serva joined our forces to give a first answer to this question in a still unpublished work\cite{nevesserva}. In that paper we devised a \textit{neutral} model able to calculate the probability distribution for the amount of Neandertal nuclear DNA in an African population as a function of the interbreeding frequency of that population with a Neandertal population. By neutral we mean that in our model Africans and Neandertals are supposed to have the same fitness.

In this work we intend to analyze the same model dropping the neutrality hypothesis. As it will be seen, the present analysis is some senses simpler than the previous\cite{nevesserva}, because we are justified in neglecting statistical fluctuations, but it introduces some technical questions rather different from the ones in our previous work. In the next section we will present the model introduced by Neves and Serva\cite{nevesserva} and give an accurate definition also of the nonneutrality hypothesis. At the same time, we will also review our previous results\cite{nevesserva} in order to compare them with the ones we will present here.

\section{The interbreeding model}

We work on the hypothesis that Neandertals and Africans lived together for some time in a restricted area, probably in the Middle East. A natural assumption then is that, due to limitation of natural resources, human population in that area remained of constant size, say equal to $N$ individuals. We will suppose that such population is divided into two subpopulations we label as 1 (Africans) and 2 (Neandertals). Moreover, we will suppose for simplicity that 
generations are non-overlapping and we will count generations from past to future. 
Reproduction is sexual and diploid and we consider random mating within each of the 
subpopulations. We also suppose that the subpopulations have lived isolated from each other 
for a long time. The time when subpopulations meet and start sharing the same environment is labeled as generation $g=0$. At that time total population consisted then of two groups, each of which composed by individuals of a pure race, which would live in the same restricted area for some time. 

The numbers $N_1(g)$ and $N_2(g)$ of individuals at generation 
$g$ in each of the two subpopulations are not supposed to be constant, although their sum $N_1(g)+N_2(g)=N$ is. In general, individuals will randomly choose mates belonging to their own subpopulation. Genetic mixing among subpopulations will be accomplished by assuming that at each generation a number $\alpha$ of random individuals from 
subpopulation 1 migrates to subpopulation 2 and vice-versa the same number of random
individuals from subpopulation 2 migrates to subpopulation 1.
In other words, $\alpha$ \textit{pairs} per generation are exchanged between subpopulations. Migrants will participate in the random mating process of their host subpopulations and contribute with their genes for the next generation just like any other individual in that subpopulation. 
Their offspring, if any, is considered as normal members of the host subpopulation. 
The parameter $\alpha$ introduced above may be non-integer and also less than 1. 
In such cases we interpret it as the average number of pairs of exchanged individuals per generation. Our hypotheses concerning exchange of individuals suggest that, although subpopulations shared the same environment, there was still some kind of cultural barrier which prevented them from freely mating. Despite that, we suppose that there was no biological barrier for interbreeding.

We will suppose that the relative fitnesses of Neandertals and Africans are respectively $1$ and $1+h$, i.e. Africans have in average $1+h$ children per each Neandertal child. Parameter $h \geq 0$ is thus the fitness difference between Africans and Neandertals. To be clear, an individual which was born among the Neandertal subpopulation, but was exchanged and had offspring among African subpopulation is considered as African for what concerns fitness. Vice-versa African born individuals living among Neandertals will have a Neandertal fitness. This hypothesis implies that fitness is considered as a cultural tract, not a genetic tract. Such an assumption is obviously disputable, but we believe it is not at all absurd. Moreover it simplifies the mathematical treatment given here and should not produce many errors for the small values of $\alpha$ and $h$ to be used here.

Let $x(g)=N_1(g)/N$ be the fraction of Africans in the population at generation $g$. We will define the \textit{history} of the subpopulation sizes as the set of values of $x(g)$ at all generations. In case the size of the population were infinite, then the subpopulation history would be determined by
\begin{eqnarray}
x(g+1)&=& \frac{(1+h)x(g)}{(1+h)x(g) \,+\, 1 (1-x(g))}  \,=\, \frac{(1+h)x(g)}{1+h x(g)}\nonumber\\
&\equiv& \label{defeta} \eta_h(x(g))  \;.
\end{eqnarray}
For finite $N$ we have statistical fluctuations and it is reasonable to suppose that the history will be specified by supposing that $N_1(g+1)$ is a binomially distributed random variable such that its expectation turns out to be $N \eta_h(x(g))$. More concretely,
\begin{equation} \label{probxg}
\mathrm{Prob}\,(N_1(g+1)=n_1) \,=\, \left(\begin{array}{c} N \\ n_1 \end{array}\right) \eta_h(x(g))^{n_1} (1- \eta_h(x(g))^{N-n_1} \;.
\end{equation}
Of course, $N_2(g+1)=N-N_1(g+1)$.

\begin{figure}
	\centering
		\includegraphics[width=0.9\textwidth]{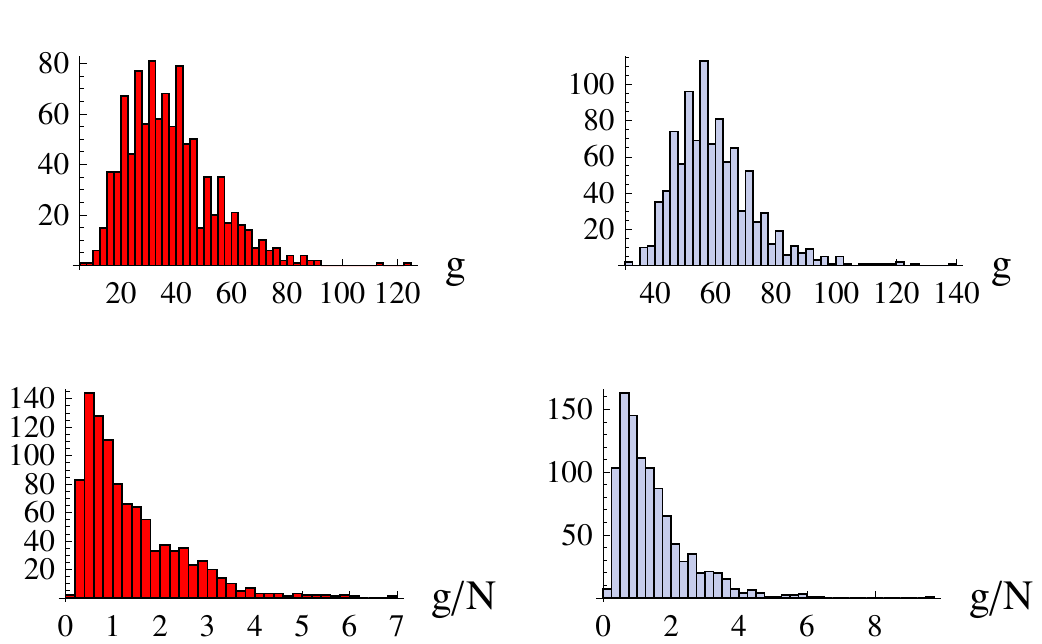}
	\caption{\label{timedists} The plots show histograms for the time to extinction empirically obtained by 1000 simulations of the stochastic processes for the population history $x$, all with $x(0)=0.5$. The two plots in the upper row are for the nonneutral case with $h=0.1$ and the two plots in the lower row are for the neutral case $h=0$. The plots in the left column refer to population size $N=100$ and the ones in the right column are for $N=1000$. Notice that whereas the horizontal axes in the plots for $h=0.1$ represent number of generations, in the plots for $h=0$ they represent number of generations divided by $N$.}
\end{figure}

In the nonneutral case $h>0$, it can be easily seen that $0$ and $1$ are the only fixed points of $\eta_h(x)$, being $0$ unstable and $1$ stable. Thus, if $N$ were infinite, $x(g) \rightarrow 1$ as $g \rightarrow \infty$. For finite $N$ statistical fluctuations ensure that $x(g)$ will be equal to 1 at some finite $g$ with very large probability if $N$ is large enough, even for small $h$. It can be shown that the mean number of generations for Neandertal extinction diverges only with $\log N$, being thus rather insensitive to $N$ in a large range of values. Moreover, for a fixed $N$, the distribution of the extinction times is relatively narrow. Such features are illustrated in Fig.  \ref{timedists}.

On the other hand, the possible set of histories in the neutral case $h=0$ is much more variable. The stochastic process for $N_1(g)$ and $N_2(g)$ turns out to be the 
neutral Wright-Fisher model for two alleles at a single locus\cite{ewens}. 
In this case, not only the time for extinction is random, but also which of the two subpopulations becomes extinct. 
Africans will be the extant subpopulation with probability $x(0)$ and the mean number of generations 
until extinction\cite{ewens} is 
\begin{equation}   \label{WFmeantime}
-2N [x(0) \ln x(0) + (1-x(0)) \ln (1-x(0))] \;.
\end{equation}
The extinction time for a fixed $x(0)$ rather than being relatively insensitive to $N$ is in fact proportional to $N$. See the illustration also in Fig. \ref{timedists}.

The expectation of $x(g+1)-x(g)$ is $\eta_h(x(g))-x(g)$ in both neutral and nonneutral cases. Whereas this quantity is exactly zero in the neutral case, it is positive in the nonneutral case. This explains the large qualitative difference between the two cases.

Our goal is to quantify the amount of Neandertal genes in Africans at the time either of Neandertal extinction or Neandertal assimilation by Africans. The technical definition of \textit{assimilation} will be explained ahead when it appears. The fraction of Neandertal genes in Africans should be compared with the experimental estimate\cite{greenetal} that living non-Africans have $1$ to $4\%$ Neandertal DNA. 

A precise definition of Neandertal and African genes follows from our hypothesis that subpopulations have been isolated from each other for a long time before $g=0$. This allows us to suppose that in many {\it loci} the two subpopulations will
have different and characteristic alleles.
Therefore, we can assume that there exists a large set of alleles which are exclusive of 
Africans and the same for Neandertals. 
We will refer to these alleles respectively 
as \textit{African} and \textit{Neandertal}.
At any generation $g\geq0$ any individual will be characterized by his/her fractions of African 
and Neandertal alleles. 
We define then $y_1(t)$ as the \textit{mean fraction of African alleles in African subpopulation
at generation $g$} and $y_2(t)$ \textit{as the mean fraction of African alleles in Neandertal subpopulation
at generation $g$}. 
The \textit{mean} here is due to the fact that individuals in African subpopulation 
in general have different allelic fractions, but $y_1(t)$ is calculated by summing 
allelic fractions of all individuals in the African subpopulation and dividing by $N_1(t)$. Similarly for $y_2(t)$. Of course at the initial generation we have
\begin{equation}   \label{initcond}
y_1(0)=1 \;\;\;\;\;\; \mathrm{and} \;\;\;\;\;\;y_2(0)=0 \;.
\end{equation} 
Similar quantities might have been defined for Neandertal alleles, 
but they are easily related to $y_1(t)$ and $y_2(t)$ and thus unnecessary.

The equations relating the mean allelic fractions at 
generation $g+1$ with the mean allelic fractions at generation $g$ were obtained in our previous work\cite{nevesserva}. In obtaining them we will further assume a \textit{mean field} hypothesis, i.e. that the $\alpha$ individuals of the African subpopulation migrating to 
the Neandertal subpopulation all have an allelic fraction equal to $y_1(t)$ and vice-versa for Neandertal migrants. The above mean field assumption is 
a strong one and it is not strictly true. Nonetheless, it is a very good 
approximation if $\alpha$ is much smaller than $1/\log_2 N$. In fact, $1/\alpha$ is the 
number of generations between two consecutive exchanges of individuals. The 
typical number of generations for genetic homogeneization in a population of $N$ individuals 
with diploid reproduction and random mating, calculated by Derrida, Manrubia and Zanette\cite{derrida1,derrida2,derrida3} and by Chang\cite{chang}, is $\log_2 N$. Thus, the condition that $\alpha$ is much 
smaller than $1/\log_2 N$ makes sure that subpopulations are both rather homogeneous 
at the exchange times. 

The allelic fraction $y_1(g+1)$ will be equal to $y_1(g)$ plus the contribution of African 
alleles due to the immigrating individuals of Neandertal subpopulation 2 and minus the loss of African
alleles due to emigration. We remind that these loss and gain terms are both proportional 
to $\alpha$ and inversely proportional to the number $N x(t)$ of individuals in the African
subpopulation. Similar considerations apply to $y_2(g+1)$. The equations below are then exactly the same as in our previous work\cite{nevesserva}, suitably rewritten in order to take into account the similarities and differences between the neutral and nonneutral cases:
\begin{equation}  \label{eqdif}
\left\{\begin{array}{rcl}
y_1(g+1) &=& \left(1- \frac{\alpha}{N x(g)}\right) 
\,y_1(g) \,+\, \frac{\alpha}{N x(g)} \, y_2(g) \\
y_2(g+1) &=& \frac{\alpha}{N (1-x(g))} \, y_1(g) \,+\,
\left(1- \frac{\alpha}{N (1-x(g))}\right) \,y_2(g) \end{array}\right. \;.
\end{equation}
It is understood that the above difference equations must be supplied with initial conditions (\ref{initcond}). 

Another condition we must impose on the above equations is that at any generation $g$ we have both
\begin{equation}   \label{positivity}
2\alpha< N x(g) \;\;\;\;\;\; \mathrm{and} \;\;\;\;\;\; 2\alpha < N(1-x(g)) \;.
\end{equation}
These conditions ensure that at generation $g$ both subpopulations have more than $2\alpha$ individuals, which implies that there exist individuals enough to be exchanged and also that after the exchange, migrants will not be the majority of any subpopulation. When the above conditions fail to hold we may consider that one of the subpopulations, even if not yet extinct, is so close to extinction that it has been assimilated by the other subpopulation. In this case, there is no more sense to consider subpopulations as separated. 

It is possible to see that the \textit{non-assimilation} conditions (\ref{positivity}) imply that 
\begin{equation}   \label{conspos}
x(g) (1-x(g)) > \frac{\alpha}{N} \;,
\end{equation}
which will be used below to  guarantee some nice mathematical properties of the solutions to (\ref{eqdif}, \ref{initcond}).

In the neutral case, due to (\ref{WFmeantime}) it is natural to define a rescaled time as $t=g/N$. As a consequence, (\ref{eqdif}) become a system of ordinary differential equations\cite{nevesserva} in the limit $N \rightarrow \infty$, 
\begin{equation}  \label{odes}
\left\{\begin{array}{rcl}
y_1'(t) &=& - \frac{\alpha}{x(t)} \, (y_1(t)-y_2(t)) \\
y_2'(t) &=& \frac{\alpha}{1-x(t)} \, (y_1(t)-y_2(t)) \end{array}\right. \;.
\end{equation}
The same rescaling does not work in the nonneutral case, because it would lead to Neandertal extinction in rescaled time equal to zero.

Another difference is that in the neutral case, as shown by (\ref{odes}), the relevant parameter for quantifying the amount of interbreeding is $\alpha$, i.e. the number of exchanged pairs of individuals per generation. In particular, larger values of $N$ imply more generations until one of the subpopulations is extinct, but the final values of the allelic fractions $y_1$ and $y_2$ are independent of $N$, depending only on $\alpha$ and on the particular realization of the history $x(t)$. On the other hand, in the nonneutral case, the relevant interbreeding parameter is 
\begin{equation}   \label{defbeta}
\beta= \frac{\alpha}{N} \;,
\end{equation}
the fraction of the entire population exchanged per generation. 

A qualitative picture of the solutions of (\ref{eqdif}, \ref{initcond}) can be obtained in a way analogous to what we did before\cite{nevesserva}. Introducing the auxiliary functions $z_1(g)=y_1(g)-y_2(g)$ and
$z_2(g)=y_1(g)+y_2(g)$, Eqs. (\ref{eqdif}) become
\begin{equation}   \label{varz1z2}
\left\{\begin{array}{rcl}
z_1(g+1)-z_1(g) &=& - \frac{\beta}{x(g)(1-x(g))} \, z_1(g) \\
z_2(g+1)-z_2(g) &=& \frac{\beta}{x(g)(1-x(g))} \, (2x(g)-1) \,z_1(g) \end{array}\right. \;.
\end{equation}

By using (\ref{conspos}), the first of these equations shows that the difference between $y_1$ and $y_2$ is always positive and decreases in time until either Neandertals are extinct or assimilated. Moreover, this difference will become close to zero if $\beta$ is large enough, or if the number of generations until Neandertal extinction or assimilation is large. If this is plugged again in (\ref{eqdif}), rewritten as
\begin{equation}
\left\{\begin{array}{rcl}
y_1(g+1)-y_1(g) &=& - \frac{\beta}{x(g)}
\,z_1(g)  \\
y_2(g+1)-y_2(g) &=& \frac{\beta}{1-x(g)} \, z_1(g) \end{array} \right. \;,
\end{equation}
we see that $y_1$ must decrease and $y_2$ must increase with time. We also see that the rates of decrease of $y_1$ and increase of $y_2$ are in general different. The transfer of Neandertal alleles to the African subpopulation and of African alleles to the Neandertals will be symmetrical only if $x(g)=1/2$. Moreover, the transfer of  African alleles to the Neandertal subpopulation and vice-versa is more effective at initial generations, when $z_1(g)$ is larger.

The above qualitative view is enough for many purposes; for quantitative purposes we may easily numerically iterate the difference equations (\ref{eqdif}), thus obtaining $y_1$ and $y_2$ at all generations. This is computationally much easier than running simulations of the entire stochastic processes of reproduction and gene transfer, described thoroughly before\cite{nevesserva}. A good agreement between results of simulations of the entire stochastic processes and solutions to (\ref{eqdif}) has been illustrated\cite{nevesserva} and will not be repeated here.

\section{The deterministic approximation and results}
Let $y_1^f$ denote the final value of $y_1$ at the time at which the Neandertal subpopulation is either extinct or assimilated. In the neutral case\cite{nevesserva} $y_1^f$ will depend on $\alpha$ and $x(0)$ and, given $x(0)$, will also depend very much on the particular realization of the history $x(g)$ which follows. As we have no information on the value of $x(0)$ and its subsequent history $x(g)$ in the particular realization of the process which produced the \textit{H. sapiens} population as it is nowadays, the best we could do in the neutral case was to produce many simulations with random values of $x(0)$, random subsequent history and random $\alpha$ and look at $y_1^f$ in such simulations. After doing that we counted the number of realizations for a particular value of $\alpha$ such that $y_1^f$ belonged to the experimental interval\cite{greenetal} of $1$ to $4\%$, i.e. $0.96 \leq y_1^f \leq 0.99$. As a result of such a process we obtained\cite{nevesserva}, in the neutral case, an empirical probability density distribution for $\alpha$ such that $y_1^f$ falls into the experimental interval. The result\cite{nevesserva} is that the maximum of this distribution is for $\alpha \approx 0.013$ and its mean value is $\alpha \approx 0.083$. As the reader may notice, by taking random histories the only parameter which needs to be estimated in the neutral case is $\alpha$. This small number of parameters to be adjusted was considered\cite{buchanan} a strong point in our previous results.

In the nonneutral case, unfortunately we have two extra parameters with respect to the neutral case: the fitness difference $h$ and the total population size $N$. On the other hand we have an advantage: the fact that the expected value of $x(g+1)-x(g)$ is positive makes the dependence on the particular history $x(g)$ much smaller than in the neutral case. As a consequence stochastic histories are very well approximated by the \textit{deterministic} history, i.e. the history obtained in the $N \rightarrow \infty$ limit, given by $x(g+1)=\eta_h(x(g))$, supplemented by the stopping condition -- without it, the history will never end --
\begin{equation}
x(g)> \min\{1-\frac{1}{N}, 1-2\beta\} \;.
\end{equation} 
The above condition means that we stop the deterministic history when either there is less than one Neandertal individual, or Neandertals were assimilated. We may then use the deterministic history in Eqs. (\ref{eqdif}) to calculate $y_1^f$. Such a calculation will be referred to as the \textit{deterministic approximation} for $y_1^f$.

\begin{figure}
	\centering
		\includegraphics[width=0.9\textwidth]{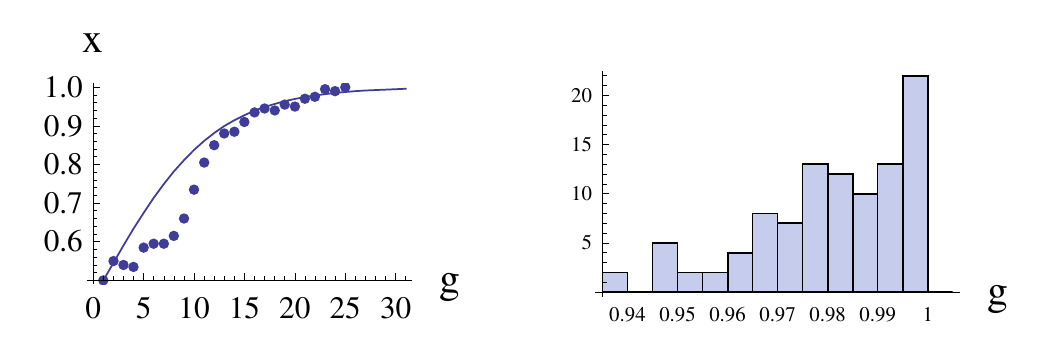}
	\caption{\label{comparisons} In the left we show plots of a simulated history (dots) and the deterministic history (full line) for a population with size $N=200$, initial fraction of Africans $x(0)=0.5$ and fitness difference $h=0.2$. In the right we show the histogram of the $y_1^f$ values obtained for 100 simulations of a population again with $N=200$, $x(0)=0.5$ and $h=0.2$. The value of the interbreeding parameter is $\beta=5 \times  10^{-4}$, which corresponds to one pair of individuals exchanged at each 10 generations. For comparison sake, the mean of the presented $y_1^f$ data is 0.980636, while the corresponding value calculated in the deterministic approximation is 0.983291.}
\end{figure}

In Fig. \ref{comparisons} we compare results from simulations of the complete stochastic processes of reproduction within subpopulations and  migrations until Neandertal extinction or assimilation with the corresponding results in the deterministic approximation. The reader may see that agreement is very good.

\begin{figure}
	\centering
		\includegraphics[width=0.9\textwidth]{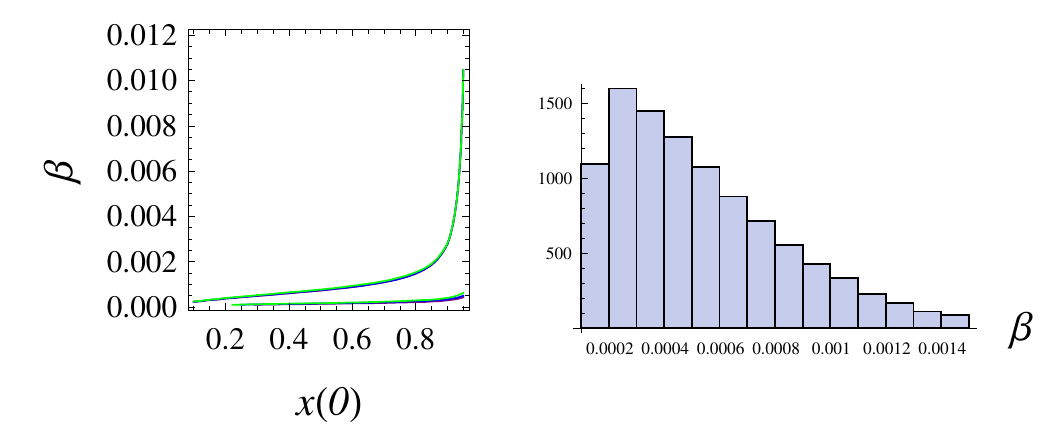}
	\caption{\label{h01} At the left we show for $h=0.1$ and values $N=100$, $N=200$ and $N=1000$ the regions in the plane $x(0)$, $\beta$ such that $y_1^f$ lies in the experimental interval between 0.96 and 0.99. The picture was obtained by calculating $y_1^f$ in the deterministic approximation for a grid of points in the plane and then drawing the level curves of $y_1^f$ at levels 0.96 and 0.99. Notice that the regions for the different values of $N$ are almost the same, mainly in the larger $\beta$ region. At the right we plot the probability density distribution for the values of $\beta$ such that $y_1^f$ in the deterministic approximation lies in the experimental interval between 0.96 and 0.99, taking also into account the no Neandertal minority assumption (\ref{80hypothesis}). The parameters are $h=0.1$ and $N=1000$. The mean value of $\beta$ is $\beta_{\mathrm{mean}} \approx 5.3 \times 10^{-4}$. The maximum is at $\beta_{\mathrm{max}} \approx 2.5 \times 10^{-4}$}.
\end{figure}

We may now proceed to our main objective, which is estimating the interbreeding parameter $\beta$ for some chosen values of the fitness difference $h$: $h=0.2$, $h=0.1$, $h=0.05$, $h=0.01$. As the left part of Fig. \ref{h01} illustrates, and as remarked before, the dependence of $y_1^f$ on $N$, all other parameters fixed, is very slight. As a consequence, we will produce results for a single value $N=1000$, which seems a reasonable value. For other values of $N$ the results for $\beta$ -- but not for $\alpha$ -- are approximately the same.

The data in Fig. \ref{h01} also illustrate the range of values of allowed values of $\beta$ for each $x(0)$. By using the data in the figure we may calculate the empirical probability density distribution for the values of $\beta$ such that $y_1^f$ lies in the experimental interval, a result analogous to our main result\cite{nevesserva} in the neutral case. This probability density distribution is plotted at the right in Fig. \ref{h01} for $h=0.1$ and $N=1000$. For this plot we have restricted the histories to those such that 
\begin{equation}  \label{80hypothesis}
x(0) \leq 0.8
\end{equation}
as in our discussion\cite{nevesserva} of the neutral case. The reason for this restriction is that larger values of $x(0)$, as can be seen in the left part of Fig. \ref{h01}, strongly shift the probability density for $\beta$ to larger values. As the mean field hypothesis included in \ref{eqdif} is not accurate for large values of $\alpha$, we add the assumption that Neandertals were at least $20 \%$ of the total population at $g=0$. Apart from the mathematical reasons\cite{nevesserva} for the \textit{no Neandertal minority assumption} (\ref{80hypothesis}), we remark that it is also reasonable from a historic point of view, since Neandertals and Africans seem to have coexisted for thousands of years in the Middle East. Of course the value $0.8$ is somewhat arbitrary, but it is the same we have chosen\cite{nevesserva} in the neutral case and is used here also for comparison sake.

In the following figures we repeat the same calculations as in Fig. \ref{h01} to the values $h=0.05$, $h=0.01$ and finally $h=0.2$. In all of them we use $N=1000$ and the no Neandertal minority assumption (\ref{80hypothesis}).

\begin{figure}
	\centering
		\includegraphics[width=0.9\textwidth]{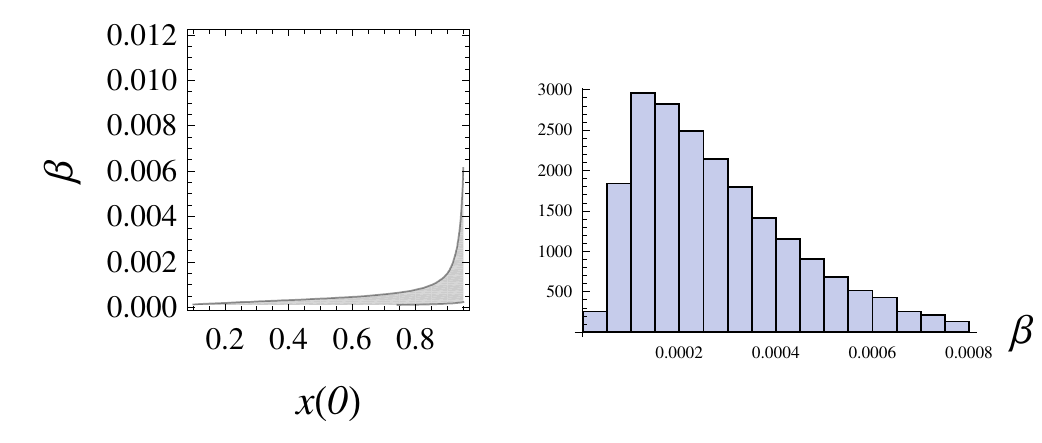}
	\caption{\label{h005} The same as in Fig. \ref{h01}, but now $h=0.05$ and we show only the region for $N=1000$. The mean value of $\beta$ is $\beta_{\mathrm{mean}} \approx 2.7 \times 10^{-4}$. The maximum is at $\beta_{\mathrm{max}} \approx 1.3 \times 10^{-4}$}.
\end{figure}

\begin{figure}
	\centering
		\includegraphics[width=0.9\textwidth]{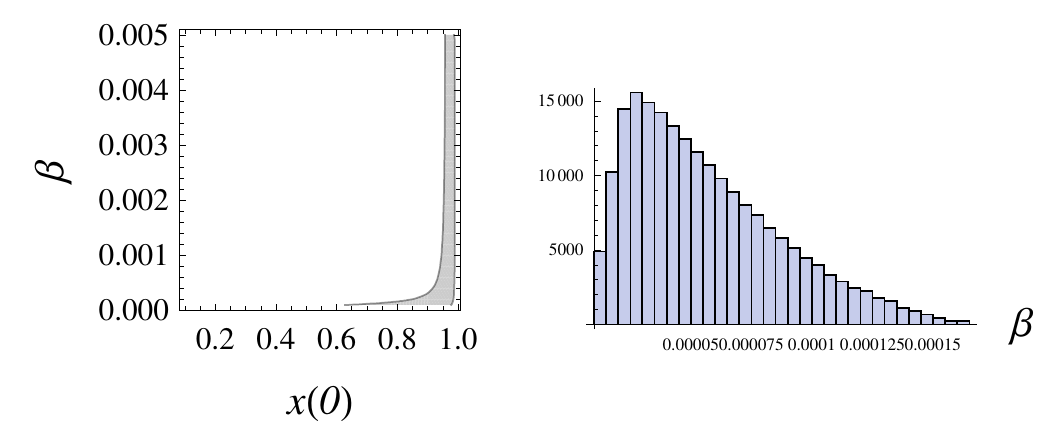}
	\caption{\label{h001} The same as in Fig. \ref{h005}, with $h=0.01$ and $N=1000$. The mean value of $\beta$ is $\beta_{\mathrm{mean}} \approx 5.4 \times 10^{-5}$. The maximum is at $\beta_{\mathrm{max}} \approx 2.8 \times 10^{-5}$.}
\end{figure}

\begin{figure}
	\centering
		\includegraphics[width=0.9\textwidth]{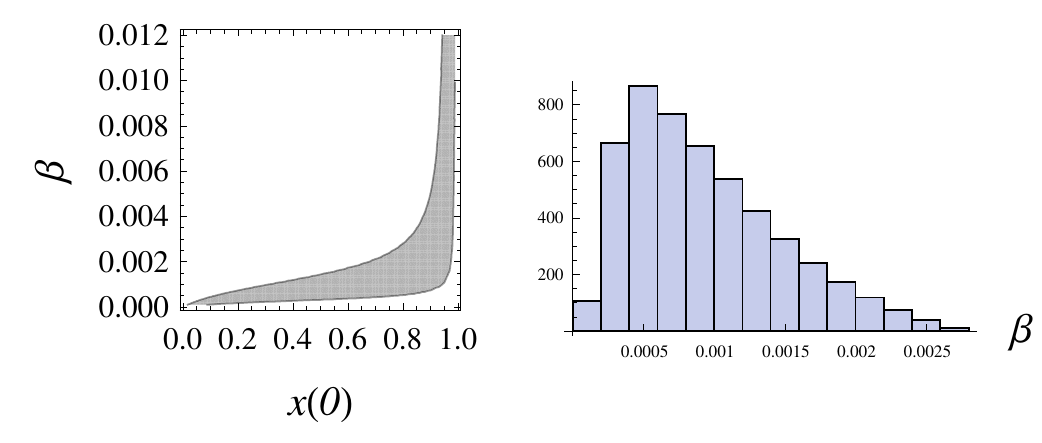}
	\caption{\label{h02} The same as in Fig. \ref{h005}, with $h=0.2$ and $N=1000$. The mean value of $\beta$ is $\beta_{\mathrm{mean}} \approx 9.3 \times 10^{-4}$. The maximum is at $\beta_{\mathrm{max}} \approx 5.0 \times 10^{-4}$.}
\end{figure}

\section{Discussion and conclusions}

We have presented here the extension to a nonneutral setting of a model\cite{nevesserva} originally intended to estimate the amount of interbreeding between anatomically modern Africans and Neandertals supposing equal fitnesses for these groups. As commented along the text there are some subtle differences in the mathematical treatment of the two cases. Due to these differences the interbreeding parameter $\alpha$ introduced in our previous work was substituted by $\beta= \alpha/N$. In doing this, results turn out to be approximately independent of $N$, as illustrated in the left part of Fig. \ref{h01}. Comparison of the present results with the previous ones\cite{nevesserva} is possible only by choosing some value for $N$. Although results in Figs. \ref{h01}, \ref{h005}, \ref{h001} and \ref{h02} were produced with the choice $N=1000$, the $N$-independence permits that we extrapolate them to other values for $N$.

\begin{figure}
	\centering
		\includegraphics[width=0.9\textwidth]{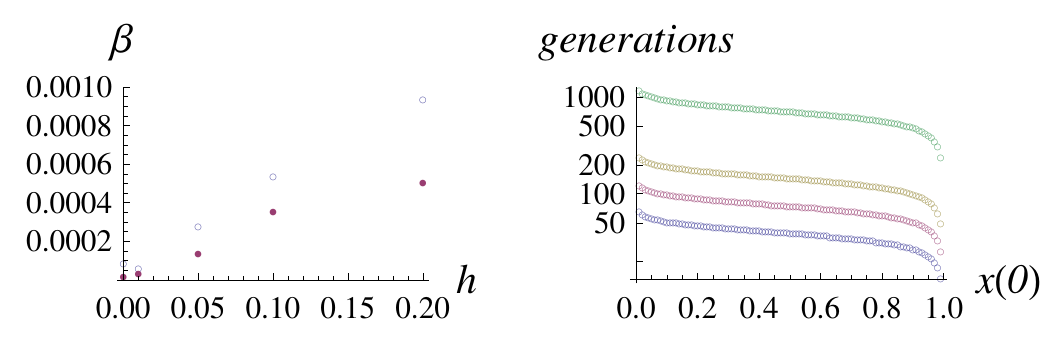}
	\caption{\label{sum} In the left picture we summarize the results of the neutral, $h=0$, and nonneutral cases, giving as functions of $h$ the positions $\beta_{\mathrm{max}}$ of the maximum of the probability density distributions and $\beta_{\mathrm{mean}}$ of their means. Filled dots are data for $\beta_{\mathrm{max}}$ and empty dots are data for $\beta_{\mathrm{mean}}$. In the right picture we show for a total population $N=1000$ the number of generations necessary for extinction of Neandertals in the deterministic approximation as functions of the initial fraction of Africans $x(0)$. The various curves correspond to the values $h=0.01$, $h=0.05$, $h=0.1$ and $h=0.2$. For fixed $x(0)$ the number of generations is a decreasing function of $h$, which makes labelling of the curves unnecessary. For other values of $N$ results do not change very much, as commented in the text.}
\end{figure}

In the left part of Fig. \ref{sum} we present a summary of the results of Figs. \ref{h01}, \ref{h005}, \ref{h001} and \ref{h02} along with the corresponding results for the neutral case $h=0$. It turns out that, as expected, larger values for the fitness difference imply shorter times until Neandertals are extinct or assimilated and, as a result, interbreeding must be more intense in order that the necessary amount of Neandertals alleles be transferred to the African subpopulation. This explains why the values of $\beta$ increase with $h$.

Although we used $\beta$ instead of $\alpha$ in this work, we remind that the mean field hypothesis used in deriving Eqs. (\ref{eqdif}) relies on the smallness of $\alpha$. More exactly, the mean field hypothesis is accurate for $\alpha \ll 1/\log_2 N$. For $N=1000$, this means $\alpha \ll 0.01$, or $\beta \ll 10^{-4}$. Many of the values found for $\beta$ in Fig. \ref{sum} are larger than that. These values should be taken with care if $N=1000$, but if $N$ were smaller, say $N=100$, they may be considered accurate. We do not know any estimate for $N$ in the Middle East, but 10,000 individuals\cite{takahata} has been suggested as a good estimate for the total human population in ancient times.

Another question to be considered in applying the present results, not yet addressed, is the time necessary for the extinction of Neandertals. We show in the right part of Fig. \ref{sum} a plot of the number of generations until Neandertal extinction, in the deterministic approximation, as a function of the initial fraction $x(0)$ of Africans for the various values for $h$ used here. Taking one generation to be roughly 20 years, we see that the times range from 600 to 1000 years for $h=0.2$ to 11,000 to 20,000 years for $h=0.01$. 

According to Bar-Yosef\cite{natgeo}, occupation of the Middle East by Neandertals and Africans can be compared with a long football game, in which teams alternated in their ability to dominate the game field. Such a situation is more likely to be described by the neutral model, but a nonneutral situation, in which Africans had a slight advantage, could be possible. Our present results show that even a 1\% fitness advantage in favor of Africans, the smallest $h$ value we investigated, seems to be too large, as the caves of Skuhl and Kafzeh, in Israel, alternated between Africans and Neandertals several times over a period of more than 130,000 years. For smaller values of $h$ it is possible that statistical fluctuations make the deterministic approximation employed here useless. As our previous work on the neutral model\cite{nevesserva} did take into account the unavoidable statistical fluctuations, we believe -- up to now -- that the neutral model is more suitable for explaining both the amount of Neandertal DNA in present day humans and the long time Africans and Neandertals lived in the Middle East.

\section*{Acknowledgements} We are grateful to Maurizio Serva for several useful discussions.



\end{document}